\documentclass[twocolumn,showpacs,preprintnumbers,amsmath,amssymb,superscriptaddress,prb]{revtex4}

\usepackage{graphicx,times}
\usepackage{dcolumn}
\usepackage{bm}

\def\U#1{{%
\def\O{\mbox{O}}
\def\u{\mbox{u}}
\mathcode`\u=\mu
\mathcode`\O=\Omega
\mathrm{#1}}}

\def\ii{{\mathrm{i}}}

\def\sub#1{_{\scriptsize\mbox{#1}}}
\newcommand{\vect}[1]{{\mbox{\boldmath $#1$}}}

\begin{document}


\title{Observation of Brewster's effect for transverse-electric
electromagnetic waves in metamaterials: Experiment and theory}

\author{Y. Tamayama}
\affiliation{Department of Electronic Science and Engineering,
Kyoto University, Kyoto 615-8510, Japan}
\author{T. Nakanishi}
\email{t-naka@kuee.kyoto-u.ac.jp}
\author{K. Sugiyama}
\author{M. Kitano}
\email{kitano@kuee.kyoto-u.ac.jp}
\affiliation{Department of Electronic Science and Engineering,
Kyoto University, Kyoto 615-8510, Japan}
\affiliation{CREST, Japan Science and Technology Agency, Tokyo
103-0028, Japan}

\date{\today}

\begin{abstract}

We have experimentally realized Brewster's effect for 
transverse-electric (TE) waves with metamaterials.
In dielectric media, Brewster's no-reflection
effect arises only for transverse-magnetic (TM) waves.
However, it has been predicted theoretically
that Brewster's effect arises for TE waves under the condition
that the relative permeability $\mu \sub{r}$ is not
equal to unity. 
We have designed an array of split ring resonators (SRRs) as a  
metamaterial with $\mu \sub{r} \neq 1$ using
a finite-difference time-domain (FDTD) method.
The reflection measurements were carried out
in a 3-GHz region and the disappearance of reflected waves
at a particular incident angle was confirmed.

\end{abstract}

\maketitle


Brewster's no-reflection condition is
one of the main features of the laws of reflection and
refraction of electromagnetic waves at a boundary between two
media.
For a specific incident angle, known as the Brewster angle,
the reflection wave vanishes.
In dielectric media, this phenomenon exists only for 
transverse-magnetic (TM) waves (p waves), 
and not for transverse-electric (TE) waves (s waves).
It is conveniently applied in optical instruments.
One can generate completely polarized light from an unpolarized
light source only with a glass plate.
It can also be used to avoid the reflection losses at
the surfaces of optical components.
The Brewster window of the discharge tube in gaseous lasers is
a typical example.

For a plane electromagnetic wave 
incident on the plane boundary between medium 1 and medium 2,
the amplitude reflectivities of TE and TM waves are given by
the Fresnel formulae:
\begin{align}
r\sub{TE} 
= \frac{\sin{(\theta \sub{t} - \theta \sub{i} )}}
{\sin{(\theta \sub{t} + \theta \sub{i} )}} 
,\quad
r\sub{TM} 
= \frac{\tan{(\theta \sub{i} - \theta \sub{t} )}}
{\tan{(\theta \sub{i} + \theta \sub{t} )}} 
,
\label{eq:1TETM}
\end{align}
where $\theta\sub{i}$ 
and $\theta\sub{t}$ are the angles of incidence 
and transmission, respectively \cite{hecht}. 
The numerators in Eq.~(\ref{eq:1TETM})
cannot vanish because $\theta \sub{i}$ is not equal to 
$\theta \sub{t}$.
However, $r\sub{TM}$ can vanish because 
$\tan{(\theta \sub{i} + \theta \sub{t} )}$ diverges to
infinity when $(\theta \sub{i} + \theta \sub{t})$ is
equal to $\pi /2$. 

Physically, Brewster's phenomena can be understood as follows.
The direction of the induced electric dipole in medium 2
is perpendicular to the wavevector therein.
With regard to TM waves, the dipole lies in the plane of incidence.
A linearly vibrating dipole radiates transversally and cannot 
emit radiation in the direction of the vibration.
This direction coincides with the wavevector of the reflected wave
when the Brewster condition is satisfied.
The oscillating dipoles in the medium 2 
do not send any waves in the direction of the reflection.
On the other hand, with regard to TE waves, each dipole is perpendicular to
the plane of incidence and emits waves isotropically in the plane.
Therefore, no special angles exist for TE waves. 
(The dipole model also explains the sign change 
in the amplitude reflectivity
when the angle is changed through the Brewster angle.)

Brewster's effect in dielectric media
exists only for TM waves, and not for TE waves.
This asymmetry results from the assumption 
that the relative permeability $\mu\sub{r}$  
is almost unity for higher frequencies, such as microwaves and light
waves.
Each medium is characterized only by its relative permittivity
$\varepsilon\sub{r}$.
The assumption that is used in deriving Eq.~(\ref{eq:1TETM}) is 
quite reasonable because for common materials, 
any kind of magnetic response is frozen in high frequency regions.
However, the assumption must be reconsidered for metamaterials,
for which both $\varepsilon\sub{r}$ and $\mu\sub{r}$ can be changed
significantly from unity. 
Brewster's effects in general cases such as
magnetic, anisotropic, and chiral madia
have been studied \cite{doyle80,lakhtakia92,futterman95}
and no-reflection phenomena for other than TM waves have been predicted.
The use of metamaterials for verifying these generalized Brewster
effects have been suggested \cite{leskova01,leskova03,fu05}.
In this Letter, we demonstrate the Brewster effect
for TE waves in magnetic metamaterials experimentally.

A metamaterial is composed of small conductive 
elements such as coils or rods, upon which
currents are induced by the incident electric or
magnetic fields.
When their sizes and separations are significantly smaller
than the wavelengths, the collection of elements 
can be viewed as a continuous medium.
By utilizing resonant structures, both $\varepsilon\sub{r}$ and
$\mu\sub{r}$ could be significantly shifted from unity.
In particular, a medium with $\varepsilon\sub{r}<0$ and
$\mu\sub{r}<0$ attracts attention because of its peculiar
behaviors owing to the negative index of refraction
$n = \sqrt{\varepsilon\sub{r}}\sqrt{\mu\sub{r}}
=-\sqrt{|\varepsilon\sub{r}\mu\sub{r}|}$ \cite{veslago}.
Negative refraction has been experimentally confirmed in 
microwave and terahertz regions \cite{shelby,parazzoili,houck}. 

An artificial medium with $\varepsilon\sub{r}=1$, 
$\mu\sub{r}\neq 1$, which is a dual of normal dielectric
materials, $\varepsilon\sub{r}\neq 1$, $\mu\sub{r}=1$,
can be designed.
Magnetic dipoles are induced by the magnetic field of the
incident wave.
Repeating the discussion with the dipole model, one can
conclude that Brewster's effect can be observed
for TE waves in the case of such magnetic metamaterials.

\begin{figure}[]
\begin{center}
\includegraphics[scale=1]{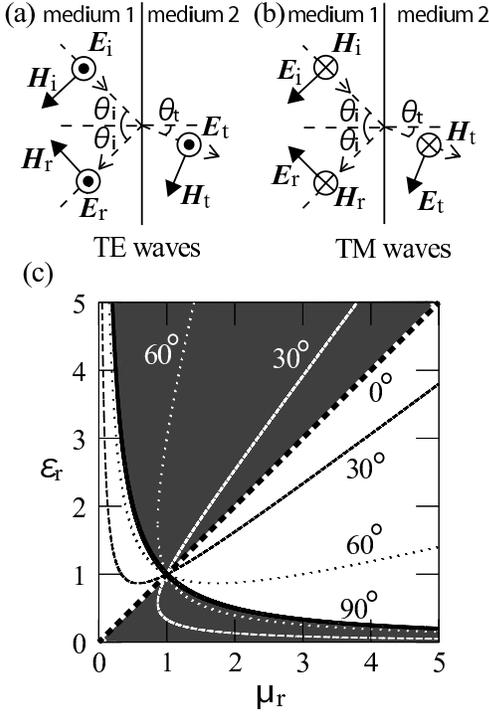}
\caption{The relation between electric and  magnetic fields for
(a) TE waves and (b) TM waves.
(c) Brewster condition for media with relative permittivity 
$\varepsilon \sub{r}$ 
and relative permeability $\mu \sub{r}$. 
Brewster conditions for TE waves and TM waves exist in the 
unshaded and shaded areas, respectively. 
The curves represent the contour lines of the Brewster angles.}
\label{fig:Brewregion}
\end{center}
\end{figure}

We assume that medium 1 is a vacuum and medium 2 is 
a medium with $\varepsilon\sub{r}$ and
$\mu _{\mathrm{r}}$.
The amplitude reflectivities for TE waves and TM waves are 
expressed as follows:
\begin{align}
r \sub{TE} 
=\frac{Z\sub{r}\cos\theta\sub{i}-\cos\theta\sub{t}}
{Z\sub{r}\cos\theta\sub{i}+\cos\theta\sub{t}}
,\quad
r \sub{TM} 
=\frac{\cos\theta\sub{i}-Z\sub{r}\cos\theta\sub{t}}
{\cos\theta\sub{i}+Z\sub{r}\cos\theta\sub{t}},
 \label{eq:RTETM} 
\end{align}
where $Z\sub{r}=\sqrt{\mu\sub{r}/\varepsilon\sub{r}}$
is the normalized wave impedance of medium 2 \cite{hecht}.
The incident angle $\theta\sub{i}$ and
the transmitted angle $\theta\sub{t}$ are related by
Snell's law
$\sin\theta\sub{i}/\sin\theta\sub{t}=n$ with
$n=\sqrt{\varepsilon\sub{r}}\sqrt{\mu\sub{r}}$.
The no-reflection conditions, $r\sub{TE}=0$ or $r\sub{TM}=0$, 
can be written as follows:
\begin{align}
  0\leq\sin^2\theta\sub{i}=
\frac{\alpha ^2-n^2}{\alpha ^2-1}\leq 1
, \label{eq:theta}
\end{align}
where $\alpha=\mu\sub{r}$ for TE waves and
$\alpha=\varepsilon\sub{r}$ for TM waves.
Theoretical derivations of the Brewster angles for general media
have been presented in previous works
\cite{lakhtakia92,futterman95,leskova01,leskova03,fu05}.
With this equation, the Brewster angle can be determined
for a given pair, $(\mu \sub{r}, \varepsilon\sub{r})$.
In Fig.~\ref{fig:Brewregion} (c), 
the Brewster angles are plotted parametrically on the
$(\mu \sub{r}, \varepsilon\sub{r})$-plane.
Based on the inequalities of Eq.~(\ref{eq:theta}), 
we see that the Brewster angle for TM waves exists only in the shaded
area in Fig.~\ref{fig:Brewregion} (c).
By exchanging the roles of $\varepsilon\sub{r}$ and $\mu\sub{r}$,
we obtain the Brewster condition for TE waves as indicated by 
the unshaded area in Fig.~\ref{fig:Brewregion} (c).

\begin{figure}[tbp]
\begin{center}
\includegraphics[scale=0.6]{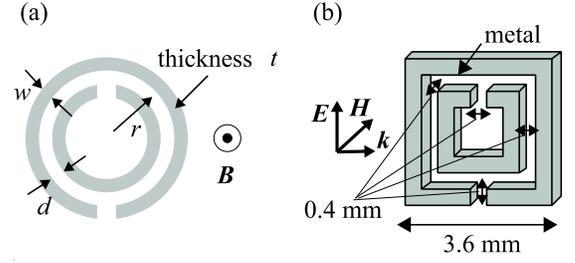}
\caption{(a) Split ring resonator. $r$ is the average 
radius of two rings;
$w$, the width of the ring;
$d$, the distance between the two rings; 
$t$, the thickness of the ring. 
(b) SRR used in the FDTD analysis. }
\label{fig:SRR}
\end{center}
\end{figure}

\begin{figure}[b]
\begin{center}
\includegraphics[scale=0.6]{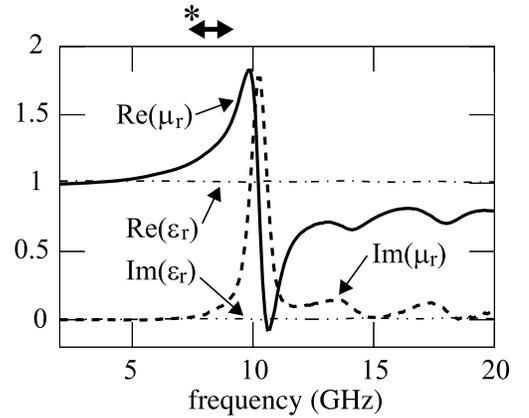}
\caption{Complex relative permittivity $\varepsilon \sub{r}$ 
and complex relative permeability $\mu \sub{r}$ versus frequency 
for the SRR array. 
The SRRs in Fig.~\ref{fig:SRR} (b) are 
placed every $4.4\,\U{mm}$ in both the $\vect{E}$ direction and 
the $\vect{H}$ direction to form a 2D array. The array is a monolayer 
in the $\vect{k}$ direction. 
The conductivity of the ring is $1.0 \times 10^8 \,\U{S/m}$. 
$R\sub{min} / R_0$ is very small in the frequency region and 
is represented by the arrow marked with an asterisk.}
\label{fig:srrepsmu}
\end{center}
\end{figure}

It is apparent that 
Brewster's effect arises only for TM waves
in normal media ($\mu\sub{r}=1$).
However, for a medium with $\mu\sub{r}\neq 1$,
the Brewster condition for TE waves can be realized.
For a given $\mu\sub{r}\gtrless 1$, when 
$1/\mu\sub{r} \lessgtr \varepsilon\sub{r} \lessgtr \mu\sub{r}$ is 
satisfied, there exists a Brewster angle for TE waves. 
In other words, the medium must be more magnetic, rather than electric,
in order to realize the Brewster condition for TE waves.


We consider an array of split ring resonators (SRRs),
as shown in Fig.~\ref{fig:SRR} (a).
It serves as a magnetic medium with $\mu\sub{r} \neq 1$ in microwave
regions \cite{srr}.
Each SRR functions as a series resonant circuit
formed by the ring inductance and inter-ring capacitance.
When we apply time-varying magnetic fields through the ring,
a circular current around the ring is
induced near the resonance frequency so that it produces a magnetic moment.
Owing to the resonant structure,
the effective permeability $\mu \sub{r}$ could be significantly
different from unity.


We calculate the relative complex permittivity $\varepsilon\sub{r}$
and permeability $\mu\sub{r}$ of the SRR array
using a finite-difference time-domain (FDTD) method \cite{fdtd}.
We consider a rectangular SRR, as shown in Fig.~\ref{fig:SRR} (b),
for the purpose of simplicity in calculation.
Assuming that a plane wave is incident on the SRR array,
we analyzed the charge distribution and the induced current,
from which the electric and magnetic dipole moments originate.
The permittivity $\varepsilon\sub{r}$ (permeability $\mu\sub{r}$)
can be derived from an electric (magnetic) dipole moment of a single SRR
and the density of the SRRs \cite{smith00}. 

Figure \ref{fig:srrepsmu} shows $\varepsilon\sub{r}$ and $\mu\sub{r}$
as a function of frequency $f$.
The real parts of $\varepsilon\sub{r}$ and $\mu\sub{r}$ are related
to the dispersive or refractive properties of metamaterials, and
the imaginary parts are related to
the losses or absorption.
As seen in Fig.~\ref{fig:srrepsmu},
the permittivity can be regarded as unity for any frequency,
and ${\rm Re}(\mu\sub{r})$ and ${\rm Im}(\mu\sub{r})$
can be approximated by the Lorentz dispersion and absorption functions. 
The ratio of the maximum value of $\left| \varepsilon \sub{r} (f) -1
\right|$ to that of $\left| \mu \sub{r} (f) -1 \right|$ is less than
$2\,\%$. 
The electric response of SRR is much weaker than the magnetic response
especially for the electric field in the direction of symmetric axis of
reflection. 
This fact agrees with previous studies \cite{gay-balmaz02,katsarakis04}.
The SRR array functions as a magnetic medium with finite losses.
We also found that the LC circuit model is helpful
in the estimation of the resonance frequency. 

\begin{figure}[t]
\begin{center}
\includegraphics[scale=0.6]{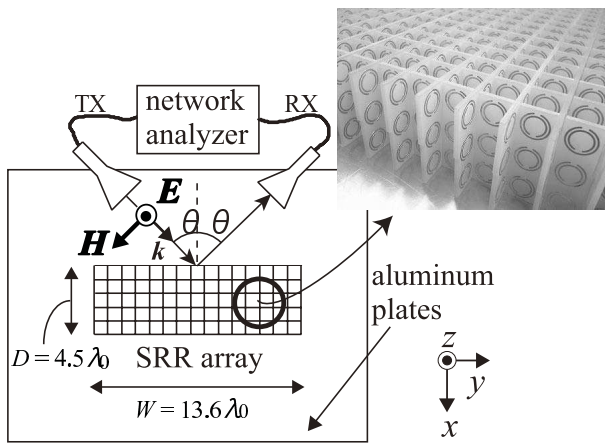}
\caption{Schematic diagram of the reflectivity measurement system. 
Measurement is performed in a 2D waveguide composed of 
two aluminum plates. 
The separation between two aluminum 
plates is $38\,\U{mm}(=0.38\lambda _0)$. 
}
\label{fig:refsystem}
\end{center}
\end{figure}

The reflectivity at the Brewster angle for the SRR array,
unlike an ideal medium without losses,
has a nonzero value due to the dissipation.
If the dispersive property of the magnetic medium dominates
the dissipation,
we can detect a significant depression in the
power reflectivity around the Brewster angle.
In order to estimate the magnitude of the depression,
we introduce the ratio of the minimum power reflectivity $R\sub{min}$
to the power reflectivity for normal incidence, $R_0$.
The lesser the ratio $R\sub{min}/R_0$,
the more easily we can detect Brewster's effect in the experiments.
From the calculation by the FDTD method,
we find that the ratio $R\sub{min}/R_0$ reduces
in a narrow region below the resonance frequency,
which is indicated by the arrow marked with an asterisk
in Fig.~\ref{fig:srrepsmu},
and the Brewster condition for the TE wave could easily be detected.


\begin{figure}[t]
\begin{center}
\includegraphics[scale=0.5]{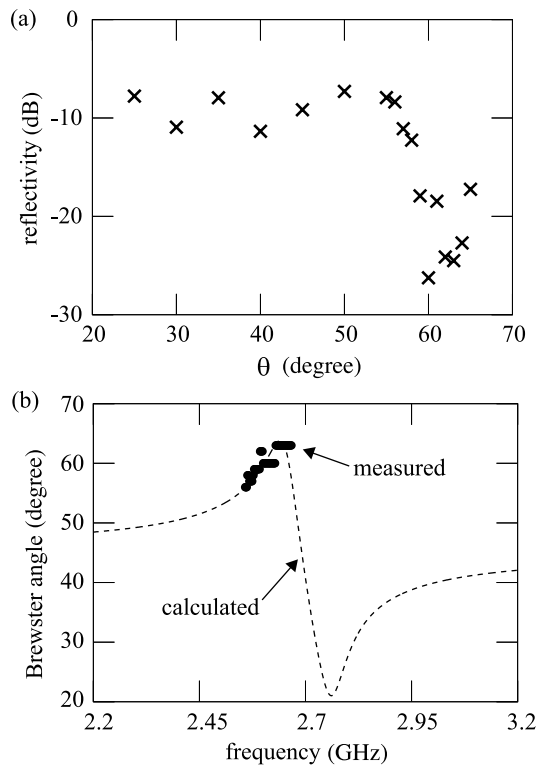}
\caption{(a) Power reflectivity of the SRR array at $f=2.6001\,\U{GHz}$ 
as a function of $\theta$. 
(b) Measured (solid circle) and calculated (dashed line) Brewster
angles versus frequency.
}
\label{fig:refbrewangle}
\end{center}
\end{figure}

A power-reflectivity measurement system as shown in 
Fig.~\ref{fig:refsystem}, is used 
to observe Brewster's effect for the TE waves with an SRR array. 

The SRRs are formed on printed circuit boards. 
To facilitate easy preparation, the parameters are chosen as 
$r=4.0\,\U{mm},~w=0.61\,\U{mm},~d=0.48\,\U{mm}$, and 
$t=35\,\U{um}$. 
From a simplified LC circuit model, 
the resonance frequency is estimated to be $3.04\,\U{GHz}$; 
this corresponds to the wavelength in a vacuum, $\lambda _0 = 
9.9\,\U{cm}$.

We set two aluminum plates ($1\,\U{m} \times 1.5\,\U{m}$) 
parallel to each other to 
form a two-dimensional waveguide, in which horn antennas and 
an SRR array are inserted. 
The separation between the two plates is $38\,\U{mm} 
(=0.38\lambda _0)$. 
In this waveguide, 
the electric field is perpendicular to the plates, and 
the electromagnetic field becomes uniform along the $z$ direction. 
Thus only TE waves can be propagated. 
We note that the direction of electric field does not change for
the change of the incident angle $\theta$ and fixed to
the direction of the minimum electric response of the SRR array.

The unit cell size of an SRR array must be significantly
smaller than $\lambda _0$; therefore,
we arranged the SRRs every $1.4\,\U{cm}(=0.14\lambda _0)$ 
in the $x$ and $y$ directions and $1.3\,\U{cm}(=0.13\lambda _0)$ 
in the $z$ direction. 
The direction of $\vect{H}$ varies in the $xy$ plane because 
the power reflectivity is measured for various incident angles $\theta$. 
We arranged the SRRs orthogonally to make the response of 
the SRR array isotropic. 
The dimension of the SRR array is $(W,D,T) = (135\,\U{cm},
45\,\U{cm},38\,\U{mm}) = (13.6\lambda _0,4.5\lambda _0,
0.38\lambda _0)$. 
In order to ensure an extended boundary, we set the width $W$ 
to be significantly larger than $\lambda _0$. 
We made the depth $D$ sufficiently large so that the influence of the 
back side reflection can be avoided.

First, we measured the transmissivity 
of the SRR array in order to determine 
the resonance frequency $f_0$,
which was found to be $2.65\,\U{GHz}$; 
it was $12.8\,\%$ smaller than the value calculated with 
the LC circuit model.   

We used a network analyzer as the microwave generator and detector. 
We connected a horn antenna (the aperture size is $15\,\U{cm} \times
3.4\,\U{cm}$) to the transmitting port of the 
network analyzer 
in order to transmit a slowly diverging beam. 
We use another horn antenna connected to the receiving 
port in order to receive the plane wave reflected at 
the boundary. 
Only the plane wave propagating normal to the antenna 
aperture can be coupled to the receiver. 
We always set the direction of the receiving horn antenna 
such that a plane wave reflected with a reflection angle equal 
to the incident angle is detected.  

We measured the $\theta$ dependence of the 
power reflectivity for a fixed frequency. 
One of the results 
is shown in Fig.~\ref{fig:refbrewangle} (a). 
Compared with the case of perfect reflection,
the reflectivity decreases by more than $27\,\U{dB}$ in the vicinity of 
$\theta = 60^{\circ}$, which corresponds to the Brewster
angle for TE waves.

We measured the frequency dependence of the Brewster angle. 
The result is shown in Fig.~\ref{fig:refbrewangle} (b) (solid circles). 
The Brewster angles could be determined only in 
a limited region just below the resonance frequency. 
In this region, the measured Brewster angles 
increase with the frequency. 
We calculated the frequency dependence of 
the Brewster angles from Eq.~(\ref{eq:RTETM}) 
by assuming $\varepsilon \sub{r} (f) =1$ 
and $\mu \sub{r} (f) = 1 - F/(f^2 + \ii \gamma f - f_0 ^2 )$,
where 
$f_0 =2.65\,\U{GHz}$ is the resonance frequency that is previously 
determined by the absorption measurement.
By fitting the calculated values
to the measured values, we fixed the parameters $F$ and $\gamma$. 
The calculated angle is shown as the dashed line 
in Fig.~\ref{fig:refbrewangle} (b). 
It increases with frequency;
this is in agreement with the experimental results.

As previously discussed,
we can observe the Brewster angles only in a limited
frequency region.
It should be noted that in actual experiments, 
the reflectivity varies somewhat erratically due 
to the interference of spurious waves or other reasons,
and the dip in reflectivity can be detected only for the cases
where $R\sub{min} / R_0$ is sufficiently small.
In the other on-resonance frequency regions, 
$R\sub{min}$ cannot be sufficiently small due to the
absorption losses.
On the other hand, $R_0$ reduces in the off-resonance regions. 

In conclusion, we observed Brewster's effect for TE waves,
which had previously never been observed for normal dielectric media. 
We need a medium whose $\mu \sub{r}$ is not equal to 
unity.
We have used a metamaterial composed of SRRs
in order to achieve a magnetic medium in a microwave region.
This is a good example of the use of metamaterials.
In terms of the parameter space $(\mu \sub{r},\varepsilon \sub{r})$, 
by introducing metamaterials, 
the rigid condition of $\varepsilon\sub{r}>0$, $\mu\sub{r}=1$
can be eliminated. 
The restrictions $\varepsilon\sub{r}>0$ and $\mu\sub{r}>0$ can also
be eliminated.
The working range of metamaterials presently
extends from microwaves to terahertz or even to
optical regions.
In the near future, we may be able to fabricate
a Brewster window for TE light.

This research was supported by the 21st Century COE Program 
No.~14213201.


\end{document}